\documentclass[10pt,conference]{IEEEtran}
\IEEEoverridecommandlockouts
\usepackage{cite}
\usepackage{amsmath,amssymb,amsfonts}
\usepackage{algorithmic}
\usepackage{graphicx}
\usepackage{textcomp}
\usepackage{xcolor}
\usepackage{hyperref}
\usepackage{multirow}
\usepackage{booktabs}
\usepackage{balance}
\usepackage{fancyvrb}
\def\BibTeX{{\rm B\kern-.05em{\sc i\kern-.025em b}\kern-.08em
    T\kern-.1667em\lower.7ex\hbox{E}\kern-.125emX}}
\begin{document}

\title{Characterization and Prediction of Questions without Accepted Answers on Stack~Overflow}

\author{\IEEEauthorblockN{Mohamad Yazdaninia\IEEEauthorrefmark{1}}
\IEEEauthorblockA{\textit{Department of CSE and IT} \\
\textit{Shiraz University}\\
Shiraz, Iran \\
yazdaninia@cse.shirazu.ac.ir}
\and
\IEEEauthorblockN{David Lo}
\IEEEauthorblockA{\textit{School of Information Systems} \\
\textit{Singapore Management University}\\
Singapore \\
davidlo@smu.edu.sg}
\and
\IEEEauthorblockN{Ashkan Sami}
\IEEEauthorblockA{\textit{Department of CSE and IT} \\
\textit{Shiraz University}\\
Shiraz, Iran \\
sami@shirazu.ac.ir}
}

\maketitle
{\let\thefootnote\relax\footnotetext{\IEEEauthorrefmark{1}{Most of the contribution of this author has been committed after he graduated from Shiraz University.}}}
\begin{abstract}
A fast and effective approach to obtain information regarding software development problems is to search them to find similar solved problems or post questions on community question answering (CQA) websites. Solving coding problems in a short time is important, so these CQAs have a considerable impact on the software development process. However, if developers do not get their expected answers, the websites will not be useful, and software development time will increase. Stack Overflow is the most popular CQA concerning programming problems. According to its rules, the only sign that shows a question poser has achieved the desired answer is the user's acceptance. In this paper, we investigate unresolved questions, without accepted answers, on Stack Overflow. The number of unresolved questions is increasing. As of August 2019, 47\% of Stack Overflow questions were unresolved. In this study, we analyze the effectiveness of various features, including some novel features, to resolve a question. We do not use the features that contain information not present at the time of asking a question, such as answers. To evaluate our features, we deploy several predictive models trained on the features of 18 million questions to predict whether a question will get an accepted answer or not. The results of this study show a significant relationship between our proposed features and getting accepted answers. Finally, we introduce an online tool that predicts whether a question will get an accepted answer or not. Currently, Stack Overflow's users do not receive any feedback on their questions before asking them, so they could carelessly ask unclear, unreadable, or inappropriately tagged questions. By using this tool, they can modify their questions and tags to check the different results of the tool and deliberately improve their questions to get accepted answers.
\end{abstract}


\section{Introduction}\label{sec:intro}
In today’s world, people use CQAs to obtain answers to their questions and benefit from experts’ knowledge. Similarly, developers use Stack Overflow to solve their problems. This website is a rich technical knowledge-sharing website that contains questions and answers concerning specific programming problems, software algorithms, coding techniques, and software development tools \cite{SO_Tour2019}. Stack Overflow depends on its users, who contribute to the community \cite{Ponzanelli2015}. Askers, who ask questions, submit their questions to receive answers, and visitors use solved problems on Stack Overflow. To ask appropriate questions, Stack Overflow helps users with hints about on/off topics, the specificity of a question, and improvement tips \cite{SO_Ask2019}. In spite of all of these, our observations show the number of questions without accepted answers is increasing on Stack Overflow. Askers mark answers that solve their coding problems as accepted answers \cite{SO_Tour2019}. We consider questions with accepted answers as resolved questions; otherwise, we name them unresolved questions. Increasing the number of unresolved questions on Stack Overflow motivated us to inquire more deeply into them.

Developers benefit from questions with accepted answers on Stack Overflow. They can ask for solutions or reuse confirmed answers. Stack Overflow contains 18 million questions, 27 million answers, 75 million comments, and 55 thousand tags, and over 50 million people visit this invaluable resource for developers each month \cite{SO_About2019}. This CQA creates a productive opportunity for programmers to ask for solutions to their coding problems, but if an asker does not get the desired answer, s/he will not benefit from the community. In fact, this community relies on getting expected answers \cite{Novielli2014}. Moreover, another approach to solve a coding problem is searching for a solution \cite{ChenC2016}. Developers will save time and effort when they find similar questions with confirmed solutions to their problems. As a result, the major upstream of Stack Overflow’s visitors include search engine websites, especially Google \cite{Amazon2019, Xia2017}. Now, it is essential that the visitors trust the answers to reuse them. According to Stack Overflow’s norms, askers will mark answers as accepted answers when the answers solve their problems \cite{SO_Tour2019}, so visitors can confidently reuse these answers. Therefore, questions with accepted answers are potentially advantageous for programmers.

\begin{figure}[htbp]
  \centering
  \includegraphics[clip, trim=4.7cm 11.3cm 4.8cm 11cm, width=.50\textwidth]{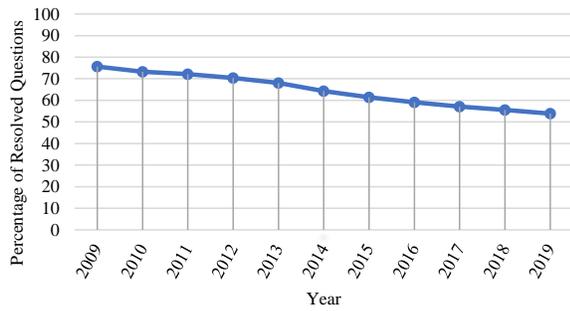}
\caption{Percentage of resolved (with an accepted answer) questions by each year, from January 2009 to January 2019}
\label{fig:trend}
\end{figure}

Our investigation into users' contributions to discussions on Stack Overflow exposed a growing problem that was increasing the number of unresolved questions, but seldom did researchers focus on it. We found 47\% of Stack Overflow’s questions had failed to get accepted answers by August 2019, while 75.7\% of the questions had been resolved by 2009 (see Figure~\ref{fig:trend}). Among over 18 million questions, 8.6 million questions have remained unresolved, while 6.6 million of these unresolved questions had received answers on Stack Overflow. Previous work mostly \cite{Anderson2012, Asaduzzaman2013, Goderie2015, Saha2013, Bhat2014, Chua2015, Baltadzhieva2015, Yao2015} focused on the questions without any answers, a subset of the unresolved questions. However, we worked on the questions with no accepted answers since questions could get useless answers. Also, prior work on Stack Overflow's questions \cite{Treude2011, Anderson2012, Asaduzzaman2013, Goderie2015, Saha2013, Bhat2014, Chua2015, Baltadzhieva2015, Yao2015, 7180109, Calefato2018, Wang2018} used sampling methods to reduce the size of data, while we accomplished a large-scale study of all Stack Overflow's data. To illustrate the importance of resolved questions, we provide three examples. The examples of answered questions without any accepted answers are presented in Figures \ref{fig:ex1}, \ref{fig:ex2}, and \ref{fig:ex3}. In Figure~\ref{fig:ex1}, the user asked for a JavaScript code snippet, but the answer was in the PHP language, so it was not useful. In the second example with well-received answers (see Figure~\ref{fig:ex2}), the asker needed a solution to add HTML comments in WordPress posts. Of course, the asker meant invisible comments, but some users suggested visible solutions. Also, others provided complex and temporary solutions. In Figure~\ref{fig:ex3}, the question was not clear enough to get the expected answer, but a user posted an answer to the question. The number of unresolved questions is increasing in the community. We, therefore, focused on such questions.

\definecolor{light-gray}{HTML}{E0E0E0}
\fboxsep=0mm
\fboxrule=0.5pt
\begin{figure}[htbp]
  \centering
\fcolorbox{light-gray}{white}{
  \includegraphics[width=.45\textwidth]{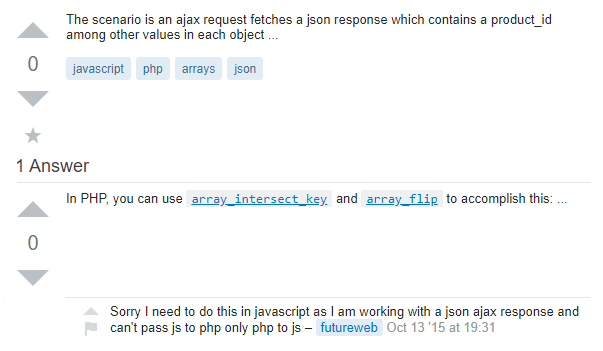}  
}
\caption{Example of an unresolved question \cite{SO2015}}
\label{fig:ex1}
\end{figure}
\begin{figure}[htbp]
  \centering
\fcolorbox{light-gray}{white}{
  \includegraphics[width=.45\textwidth]{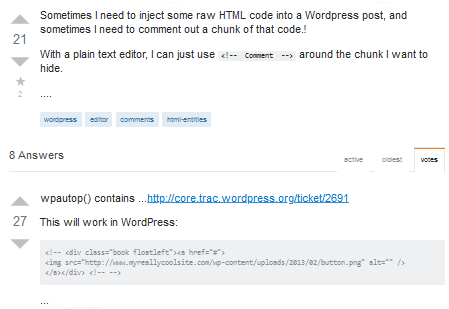}
}
  \caption{Example of an unresolved question \cite{SO2010}}
\label{fig:ex2}
\end{figure}
\begin{figure}[htbp]
  \centering
\fcolorbox{light-gray}{white}{
  \includegraphics[width=.45\textwidth]{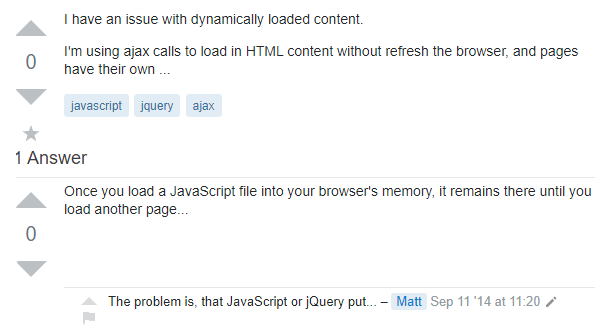}
}
  \caption{Example of an unresolved question \cite{SO2014}}
  \label{fig:ex3}
\end{figure}

In this study, we design new features that can affect getting accepted answers and propose an online tool to help Stack Overflow users by predicting whether a question will get an accepted answer or not. Toward this goal, we processed all the questions and designed several important features. To reveal the role of the tags of questions, we deeply investigated tags and introduced several novel tag-related features that can be considered as measurements to compare programming topics, such as programming languages. On the other hand, we prevented some features that could bias our results. As clarified, we focused on the features of questions, so these features should not be extracted after submitting them. For example, the features of received answers to the questions are not available at the time the questions are posted. Furthermore, in Stack Overflow data dump\footnote{\url{https://archive.org/download/stackexchange}}, some features were calculated at the time the data was released, such as Reputation. Thus, we eliminated the features that directly contain any information that is not available at the time the questions are submitted, such as view count, Reputation, and the features of answers to the questions. We evaluated our proposed features with predictive models trained on the features of 18 million questions. Finally, we introduced an online tool to help Stack Overflow users by predicting whether a question will get an accepted answer or not. The results of this study will help us to answer the following research questions in Section~\ref{sec:res}, and we will discuss how askers can receive more accepted answers in Section~\ref{sec:disc}.

\begin{itemize}
\item \textbf{RQ1}: To what extent, can we reliably predict whether a question will receive an accepted answer from its features?
\item \textbf{RQ2}: What are the most important features that can help to predict if a question will receive an accepted answer?
\item \textbf{RQ3}: Do reading hints, rules, and information about asking questions on Stack Overflow affect getting accepted
answers?
\item \textbf{RQ4}: What roles do programming topics, tags, play in resolving questions?
\end{itemize}

In the following, related work is described in Section~\ref{sec:rel}. Section~\ref{sec:back} explains the Stack Overflow website and our data collection. Then, in Section~\ref{sec:method}, we design our features. Next, we present our results in Section~\ref{sec:res} and discuss how can users get more accepted answers in Section~\ref{sec:disc}. We mention threats to validity in Section~\ref{sec:threats}. Finally, Section~\ref{sec:conc} concludes the paper.

\section{Related Work}\label{sec:rel}
Questions posted on Stack Overflow are notably of scholarly interest. Among related studies, most of them have focused on Stack Overflow’s questions with any answers, not the accepted answers. In this section, we discuss studies that have worked on questions without any answers and with accepted answers.

Some researchers focused on the questions without any answers, accepted or not. Ashton et al. \cite{Anderson2012} used machine learning algorithms to predict long-lasting questions concerning getting any answers. In the same line of research, Asaduzzaman et al. \cite{Asaduzzaman2013} analyzed 400 questions without answers on Stack Overflow. They classified the questions based on their analysis and built a classifier with the precision of 63.6\% to predict how long will a question remain unanswered. Likewise, Goderie et al. \cite{Goderie2015} built a model trained on 160K of Stack Overflow's questions to predict answering time, based on tag-related features. Saha et al. \cite{Saha2013} also worked on unanswered questions. They deployed several classifiers learned with a sample of 600K questions. As a result, questions’ scores and views were the most important factors in their model. Bhat et al.\cite{Bhat2014} were interested in question response time, receiving answers, so they built learning models to predict the response time of questions. They considered tags as significant factors in question response time. Chua and Banerjee \cite{Chua2015} were interested in ``\verb|Java|'' tagged questions that had failed to attract any answers. They presented a prediction framework learned with manually extracted features from 3000 sampled questions. In another work by Baltadzhieva and Chrupala \cite{Baltadzhieva2015}, question quality was investigated by assessing questions’ features, including questions’ scores and the number of received answers. Moreover, Yao et al. \cite{Yao2015} investigated questions with high votes on Stack Overflow and Math Stack Exchange. They considered high-score questions and answers as high-quality posts and deployed a co-prediction classifier to predict them.

\begin{figure}[htbp]
  \centering
  \includegraphics[width=.43\textwidth]{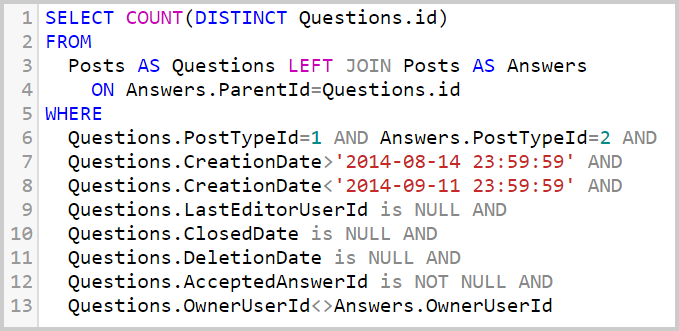}
\caption{T-SQL code to count the number of resolved questions posted from August 15, 2014, to September 11, 2014}
\label{fig:tsqlres}
\end{figure}

Dealing with resolved questions, Treude et al. \cite{Treude2011} defined different categories of questions, such as how-to or asking for an opinion. They provided the percentages of 385 questions, including unanswered, with unaccepted answers, and with accepted answers, per category, and concluded the categories can affect getting more answers or accepted answers. Rahman and Roy \cite{7180109} worked on 8000 questions that had received at least 10 answers. They extracted their proposed features, such as Reputation and votes, after submitting questions and receiving more than 9 answers. These authors built a model to assess four factors that indicates the causes of remaining unresolved questions, which had received more than 9 answers, different from our topic. We also work on all 18 million questions on Stack Overflow. Calefato et al. \cite{Calefato2018} worked on 87K questions on Stack Overflow. They extracted several categorical features to resolve a question and validated them with a model with an AUC of 0.65. Their features were classified into four classes, including sentiment, time, presentation quality, and reputation. As a result, they introduced presentation quality as the most relevant factor to resolve questions. They worked on the questions that had been asked in a month, from August 15, 2014, to September 11, 2014. During doing their work, 35\% of their dataset were with accepted answers. Due to getting new answers or acceptance marks after that period, our analysis showed that 59\% of the questions posted in that month have received accepted answers, as calculated with their preferences using Stack Exchange data explorer\footnote{\url{https://data.stackexchange.com/}} (see Figure \ref{fig:tsqlres}). The numbers of resolved questions and total questions have changed from 31K and 87K to 34K and 57K, respectively, after that month. Therefore, it shows more than 10\% of their sample of questions received accepted answers afterward. Besides, 34\% of the questions have been deleted after that month while they filtered deleted questions. On the other hand, we work on 18 million questions on Stack Overflow. We also deeply focus on programming topics, tags, and prevented the features of questions that are available after submitting them.
In another work on question response time, Wang et al. \cite{Wang2018} built a model to predict the response time to get an accepted answer to a question with more than one score that is different from our goal. They worked on a sample of 55K Stack Overflow's questions that were asked in 2015. With respect to their goals, they used some features that were not available at the time of submitting questions, such as features from answers to the questions and the owner of answers.

Different from the cited work, we investigate 18 million unresolved questions on Stack Overflow, and we do not use the features that contain information extracted after the time of posting questions. To the best of our knowledge, all related work used sampling methods, while we do not use any sampling method in this paper. We also do not filter questions without any scores or a number of answers. Furthermore, we propose several novel features that affect getting accepted answers. To reveal the role of tags in this issue, our new proposed features include several novel tag-related features. Additionally, to make our evaluation method unbiased with reference to our goal, we design features that do not contain information obtained after the time of submitting the questions, such as Reputation and answers. These features will mislead the results of our evaluation if they are extracted after submitting questions. For example, the features of answers and getting acceptance marks are not independent, so this dependency can falsely boost prediction models. This approach also creates an opportunity for us to develop a tool\footnote{\url{http://yazdaninia.info/unresolved-questions/qp/} (anonymized)} to predict whether a question will get an accepted answer or not before posting it.

\section{Background}\label{sec:back}
Stack Exchange officially publishes Stack Overflow's data as XML files. The latest set at the time of performing this study contains data from July 2008 to August 2019. During this research, we used all posts, tags, badges, and users in the data dump. These entities form a rational database, so we imported the XML files to a Microsoft SQL Server database to facilitate our feature extraction process. Due to the large size of the dataset, we ran our analysis on two powerful virtual computers, including an instance of \verb|ml.p3.16xlarge| (Amazon SageMaker ML Notebook with 64 processing cores of Intel Broadwell, 8 graphics processing units of NVIDIA Tesla V100, 488 GB of memory, and 128 GB of GPU Memory) and an instance of \verb|z1d.12xlarge| (Amazon EC2 with 48 virtual processing cores and 384 GB of memory). To prepare the data for feature design, we first analyzed the Stack Overflow website and preprocessed the data.

\subsection{Stack Overflow Website}\label{subsec:so}
The data of posts and users are the main source of our study, but there are other valuable records in the data dump. Badges, scores, and tags contain noteworthy information in terms of our goal. Stack Overflow community motivates users with a reputation system to engage them \cite{Sasso2017, Vasilescu2013}. Users can earn badges based on their activities \cite{SO_Badges2020}. For instance, ``\verb|Scholar|'' badge shows whether a user asked a question that received an accepted answer or not. ``\verb|Informed|'' badge indicates that the owner has read Stack Overflow’s tour page \footnote{\url{https://stackoverflow.com/tour}}, which provides hints on asking high-quality questions. In addition to the badges, users can increase their Reputations by participating in discussions \cite{SO_Rep2020}. For example, when answers solve askers' problems, they will mark the answers as accepted answers, and their Reputation scores will increase by 2 points \cite{SO_Tour2019, SO_Rep2020}. Also, when a question receives an upvote from users, the asker’s Reputation score will increase by 10 points. Another important part of data is tags. Tags indicate the topic of the questions. All questions are tagged in a range of 1 to 5, and users can choose their favorite tags, as expertise, to subscribe to questions with the selected tags. With respect to attracting experts, tags are essentially important. Users, in fact, follow tags, so each tag has a number of subscribers who answer questions labeled with it.

\subsection{Data Collection}\label{subsec:data}
As clarified earlier, badges, tags, users, and posts include the information that can help us to achieve our goals. Stack Overflow declares more than 100 different badges. Among these badges, we selected relevant, popular badges that could affect getting accepted answers. The badges have been awarded about 300 to 5.5 million times \cite{SO_Badges2020}, so collecting unpopular badges will lead a sparse feature set. Therefore, we collected the badges that could be related to users' knowledge and experiences in asking and answering questions (see Table~\ref{tab:features}) and had been awarded more than 30K times. Next, we assigned earned badges prior to submitting questions to each question owner.

We extracted a collection of previous contributions of users, including the history of questions and answers that had been posted by each question poser before asking questions. We also preprocessed the dataset to improve its quality and prepare it for feature extraction and building predictive models. To be sure that questions had the chance of getting accepted answers, we removed posted questions in the last 15 days in the published dataset; 90.01\% of questions receive answers within one day \cite{Bhat2014}. Also, our ran queries revealed 95.23\% of questions receive accepted answers within 15 days. Questions consist of text, code, and HTML tags. After extracting HTML and code-based features, we filtered questions to remove HTML tags and code. This led us to extract a part of content-related features. Another interesting data to collect was tags. Tags indicate the scope of questions and the number of users following them; however, the data misses and/or contains inaccurate values of the creation dates of the tags and the numbers of their followers. Thus, we developed code to fetch the numbers of tags' followers from Stack Overflow website. Also, we proposed an estimation metric of the creation dates of tags. On account of the ordered insertion of the tags into the database, we introduced the following metric, which we named it ``\verb|time_index|''. It equals to the scaled value of tag ID, the primary key of tags’ table, divided by the period value (the maximum distance of tag IDs). As presented below, we calculated the logarithm (with the base of 10) of the result to compress its range. For example, \(log(\frac{133634}{(139961-1)}*1.0-e7)) = 6.98\).

\begin{equation}
\begin{aligned}
time\_index=\log(\frac{tag\_id}{period}*\alpha)\;\;\;\;\;\;\;\;\;\;\;\;\\
   \alpha:scaling factor, period=Max(ID)-Min(ID)\;
\label{eq:ti}
\end{aligned}
\end{equation}

\section{Method}\label{sec:method}
This section presents our extracted features, which can affect attracting accepted answers, and our method to evaluate them. We extracted relevant features that were used in the prior studies. Besides, we proposed new features, including several novel tag-related features. Our extracted features and developed code are available online\footnote{\url{http://yazdaninia.info/unresolved-questions/rpd} (anonymized)}.

\begin{table*}[htbp]
\caption{Extracted features for 18 million questions calculated at the time of submitting to the community}  \label{tab:fd}
  \begin{center}
\begin{tabular}{|c|l|p{7.5cm}c|}
   \hline
    \textbf{Category}&\textbf{Title}&\textbf{Rationale}&\textbf{In Related Work}\\
    \hline
\multirow{17}{4em}{\textbf{\textit{Content}}}&Title AVG Word in Characters&Presentation quality of the questions,
readability&\cite{Kitzie2013}\\
    &Has Wh-word in Title&Clear or easy to answer &\cite{Li2012}\\   
    &Body AVG Word in Characters&Presentation quality of a question, readability&\cite{Kitzie2013}\\
    &Body AVG Sentence in Words&Presentation quality of a question, readability&\cite{Li2012}\\
    &Body Word Count&Presentation quality of a question, readability&\cite{Chua2015, Yang2011, Li2012, Treude2011}\\
    &Number of Links&Presentation quality of a question&\cite{Wang2018}\\
    &Number of Code Snippets&Presentation quality of a question, shows more details&\cite{Chua2015}\\
    &Title Word Count&Presentation quality of a question, readability&\cite{Chua2015, Yang2011, Li2012}\\
    &Title Starts with a Capital Character&Presentation quality of a question, readability&\cite{Wang2018}\\
    &Number of Paragraphs&Presentation quality of a question, readability&{New}\\
    &Is Title an Interrogative Sentence&Clear or easy to answer &{New}\\
    &``\verb|error|''/``\verb|not working|'' in Title&Clear or easy to answer &{New}\\
    &Has Quote (``\verb|<quote>|'' in Body)&Presentation quality of a question, shows more details&{New}\\
    &Lines of Code&Presentation quality of a question, shows more details&{New}\\
    &Body Sentence Count&Presentation quality of a question, readability&{New}\\
    &Code Snippets Length in Characters&Presentation quality of a question, shows more details&{New}\\
    &Has List (``\verb|<li>|'' in Body)&Presentation quality of a question, readability&{New}\\
    \hline
\multirow{6}{4em}{\textbf{\textit{Tag}}}&Tag Count&Attract more experts&\cite{Wang2018, Calefato2018, Chua2015}\\
    &Max Tag Quality&The maximum quality of a question's topic&{New}\\
    &AVG Tag Quality&The average quality of a question's topic&{New}\\
    &Max Expert Ratio&The number of experts in each topic to answer a question&{New}\\
    &Min Tag Quality&The minimum quality of a question's topic&{New}\\
    &Max Problem Rate&The maximum rate of producing questions with similar topic&{New}\\
    \hline
\multirow{3}{4em}{\textbf{\textit{Metadata}}}&Asking Time (Day of the Week)&May affect the number of visitors, experts&\cite{Asaduzzaman2013, Calefato2018, Chua2015, Treude2011}\\
    &Asking Time (Hour)&May affect the number of visitors, experts&\cite{Calefato2018, Chua2015, Li2012, Yang2011, Treude2011}\\
    &Question Creation Date&Number of unresolved questions may varies each year&{New}\\
    \hline
\multirow{26}{4em}{\textbf{\textit{User}}}&Membership Duration&Long-term users may get more accepted answers&\cite{Chua2015}\\    
    &Number of Asker's Answers&Potentials of an asker in the past&\cite{Asaduzzaman2013, Chua2015, Li2012}\\
    &Number of Asker's Questions&Potentials of an asker in the past&\cite{Asaduzzaman2013, Chua2015, Saha2013, Yao2015}\\
    &Number of Asker's Accepted Answers&Potentials of an asker in the past&\cite{Asaduzzaman2013}\\
    &Sum of Asker's Answers' Scores&Potentials of an asker in the past&{New}\\
    &Sum of Asker's Questions' Scores&Potentials of an asker in the past&{New}\\
    &Earned Scholar Badge&Marked an answer as accepted in the past&{New}\\
    &Earned Tumbleweed Badge&Had a question with no reply and score for a week&{New}\\
    &Earned Informed Badge&Read tour page \cite{SO_Tour2019} in the past&{New}\\
    &Earned Autobiographer Badge&Completed about me on Stack Overflow in the past&{New}\\
    &Earned Student Badge&Got 1+ score for the first question in the past&{New}\\
    &Earned Supporter Badge&Did an up vote in the past&{New}\\
    &Earned Editor Badge&Did an edit in the past&{New}\\
    &Earned Commentator Badge&Left 10+ comments in the past&{New}\\
    &Earned Teacher Badge&Got 1+ score for an answer in the past&{New}\\
    &Earned Analytical Badge&Visited all section of FAQ on Stack Overflow in the past&{New}\\
    &Earned Popular Question Badge&Posted a question with 1K+ views in the past&{New}\\
    &Earned Enthusiast Badge&Visit the community every month in the past&{New}\\
    &Earned Custodian Badge&Did a review task in the past&{New}\\
    &Earned Good Answer Badge&Posted an answer with the score of 25+ in the past &{New}\\
    &Earned Famous Question Badge&Posted a question with 10K+ views in the past&{New}\\
    &Earned Curious Badge&Posted a well-received question on 5 days in the past&{New}\\
    &Earned Nice Answer Badge&	Posted a question with the score of 10+ in the past&{New}\\
    &Earned Yearling Badge&Earned 200 reputation on community during a year in the past&{New}\\
    &Earned Necromancer Badge&Posted an answer after 2 months with the score of 5+ in the past&{New}\\
    &Earned Notable Question Badge&Posted a question with 2.5K+ views in the past&{New}\\
  \hline
  \end{tabular}
  \label{tab:features}
\end{center}
\end{table*}

\subsection{Feature Design}\label{subsec:feature}
To construct an insight into the features of unresolved questions, we investigated 35 questions without any answers, with unaccepted answers, or with accepted answers. We started with 25 randomly selected questions to design our features. Next, we checked 10 more randomly selected questions to extract more features, but we could not add more features to our feature set. In some cases, questions were readable and included details, but posted answers to them did not solve the problems, such as in Figure~\ref{fig:ex2}. In another scenario, questions were not clear enough to receive expected answers, such as in Figure~\ref{fig:ex3}. Posting useless answers were caused by a misunderstanding about the topic and doubt, such as in Figure~\ref{fig:ex1}. The high level of difficulty, vagueness, misunderstanding, and unpopular topics were features of these questions. In other types of unresolved questions, the lack of attraction for an expert member, incomprehensibility, duplication, and difficulty level were the main reasons for not answering the questions. Our feature design was based on this analysis. We also extracted some features cited in the prior work for 18 million questions. To reveal the importance of these features, we will evaluate them using machine learning techniques in Section~\ref{subsec:feval}.

\subsubsection{Prior Features}\label{subsubsec:pfeat}
The mentioned related work yielded various features based on their goals, including extracted features from questions’ body, title, tags, and metadata, users’ previous activities, and future answers and comments. Among these features, we used a part of them that was beneficial with respect to our objectives and immediately available at the time a question is submitted. These features included registration time \cite{Chua2015}, the average length of words in the question’s body and title \cite{Kitzie2013}, the average length of sentences in the question’s body \cite{Li2012}, the number of words in the question’s body \cite{Chua2015, Yang2011, Li2012, Treude2011}, the number of previous questions posted by the question owner \cite{Asaduzzaman2013, Chua2015, Saha2013, Yao2015}, the number of answers posted by the question owner \cite{Asaduzzaman2013, Chua2015, Li2012}, the number of answers with accepted marks posted by the question owner \cite{Asaduzzaman2013}, the number of code snippets in the question \cite{Chua2015}, asking time (hour) \cite{Calefato2018, Chua2015, Li2012, Yang2011, Treude2011}, asking time (the day of the week) \cite{Asaduzzaman2013, Calefato2018, Chua2015, Treude2011}, word count in the title \cite{Chua2015, Yang2011, Li2012}, tags count \cite{Wang2018, Calefato2018, Chua2015}, links count in the question \cite{Wang2018}, whether the title starts with a capital letter or not \cite{Wang2018}, and whether the title is a wh-question or not \cite{Li2012}. To achieve a better result, we also added more important features that show more about users’ backgrounds, such as earned badges, and novel features to reveal the role of tags in this study.

\subsubsection{New Features}\label{subsubsec:nfeat}
The efforts in Section~\ref{subsec:data} allowed us to design novel features (see Table~\ref{tab:features}). We fetched earned badges by askers for each question at the time of submitting their questions. These earned badges show users’ participants, knowledge, and experiences in posting questions and answers \cite{SO_Badges2020}. For instance, \verb|Scholar| indicates its owner has accepted at least one answer, while \verb|Tumbleweed| shows its owner asked a question that has remained unanswered, with a score of zero, without comments, and a few visitors for at least a week. In addition to the badges, we extracted some content-related factors concerning presentation quality. Some features could represent readability, including the number of paragraphs and sentences and using some HTML tags, such as \verb|<li>|, in questions. Providing details also could be shown with lines of code (LOC), code length, and using \verb|<quote>|. Some keywords, such as ``\verb|error|'' and ``\verb|not working|'', or interrogative sentences could be the signs of straightforward questions. Furthermore, earned scores by askers before submitting their questions reflect their participants and potentialities, so we calculated the sum of askers' scores earned by asking or answering for each question. To evaluate the role of tags in getting accepted answers, we also designed several novel features. Our goal was not only to design for performance but also to present meaningful factors to achieve question resolution. In Table~\ref{tab:nfeatures}, ``\verb|count|'' represents the number of questions for each tag. ``\verb|Popularity|'' quantifies tag’s popularity, which means the number of followers with respect to its creation date, ``\verb|time_index|'' \eqref{eq:ti}. Also, ``\verb|Expert Ratio|'' is the number of followers for each tag. To quantify the number of asked questions with a specific tag concerning its creation date ``\verb|Problem Rate|'' is introduced. Since questions with a popular tag would be posted at a higher rate than an unpopular tag, we divided popularity by the count of tags. We consider it ``\verb|Tag Quality|''. Due to the correlations between the presented formulas and the different numbers of tags in a question, in a range of 1 to 5, we calculated the average, minimum, and maximum of the tag-related factors of assigned tags in a question and selected a combination of five meaningful factors to append to our feature set (see Table~\ref{tab:features}).

\begin{table}
  \caption{Novel tag-related features}
  \label{tab:nf}
\begin{center}
\renewcommand{\arraystretch}{1.9}
  \begin{tabular}{|l|l|c|}
    \hline
    \textbf{Name}&\textbf{Description}&\textbf{Formula}\\
    \hline
    Popularity&Popularity of a tag&$\frac{\textstyle \#followers}{{\textstyle time\_index}}$\\
    Expert Ratio&Number of followers per tag&$\frac{{\textstyle \#followers}}{{\textstyle count}}$\\
    Problem Rate&Asking rate\footnotemark of a tag&$\frac{{\textstyle count}}{{\textstyle time\_index}}$\\
    Tag Quality&Asking rate per popularity&$\frac{{\textstyle \#followers}}{{\textstyle time\_index*count}}$\\
  \hline
\end{tabular}
  \label{tab:nfeatures}
\end{center}
\end{table}
\footnotetext{Number of asked questions with respect to time, tag's creation date}

\subsection{Features Evaluation}\label{subsec:feval}
To evaluate our proposed features with respect to getting accepted answers, we trained several classifiers on 18 million records. 47\% of the 18 million questions were without accepted answers and the models predict whether a question will receive an accepted answer or not. We also trained the models on prior features used in the related work and our new features, separately. This led us to discuss the importance of our new features in Section~\ref{subsec:rq2}. To validate the trained models, 10-fold cross-validation was used. This method prevents overfitting \cite{friedman2001elements}. We also measured the performance of the classifiers using AUC, Area Under the Curve (receiver operating characteristic). The range of AUC is $[0, 1]$; a random classifier has an AUC of 0.5, larger values of AUC show more informative classifiers, and a perfect classifier performs with an AUC of 1 \cite{han2011data}. Using threshold-independent measures, AUC, is preferable to threshold-dependents, such as sensitivity and F-score \cite{Viviani2019, Kla2018}. AUC is independent of classification threshold, assesses both costs and benefits, and is recommended to compare classifiers \cite{Bradley1997, han2011data, Provost1998}. Based on the mentioned setup, we identified important features that could affect getting accepted answers.

\section{Results}\label{sec:res}
In this section, we present our results and provide answers to RQs 1, 2, 3, and 4.

\subsection{RQ1. To what extent, can we reliably predict whether a question will receive an accepted answer from its features?}\label{subsec:rq1}
The results of the trained models reveal the relationship between the features of a question and getting an accepted answer. Table~\ref{tab:auc1} contains the performance of the classifiers that performed with AUC results of 0.60 and more. As shown in Table~\ref{tab:auc1}, the designed features resulted in considerable performance. The trained models on new features also show notable results. Classifiers with AUC scores of 0.70 and more have an adequate discrimination ability \cite{hosmer2013applied, Romano2011, Lessmann2008}. In the results, the XGBoost algorithm \cite{Chen2016} (with $max\_depth: 20$, $min\_child\_weight: 1$, $\gamma: 15$, $\eta: 0.56$,  $colsample\_by\_tree: 0.5$, $num\_parallel\_tree: 8$) attained the best performance with an AUC of 0.71. The XGBoost classifier trained on new features also performed with an AUC of 0.70, while this score for prior features was 0.66. Thus, the AUC scores demonstrate a considerable relationship between the features of questions and getting expected answers. Therefore, we can adequately predict whether a question will get an accepted answer or not from its features with an AUC of 0.71.

\begin{table}
  \caption{AUC scores of trained classifiers on all, prior, and new features}
\begin{center}
  \begin{tabular}{|l|ccc|}
    \hline
   \multicolumn{1}{|l}{\textbf{Algorithm}}&\multicolumn{3}{|c|}{\textbf{Trained on}} \\
  \cline{2-4} 
   \multicolumn{1}{|l|}{\textbf{Name}}&\textbf{All Features}&\textbf{New Features}&\textbf{Prior Features}\\
    \hline
   XGBoost&0.71&0.70&0.66\\
   CART&0.68&0.67&0.64\\
   Bayesian Ridge&0.68&0.67&0.56\\
   Ridge&0.68&0.67&0.56\\
   Lasso&0.62&0.62&0.51\\
   GNB$^{\mathrm{a}}$&0.61&0.61&0.55\\
  \hline
  \multicolumn{4}{l}{$^{\mathrm{a}}$Gaussian Naive Bayes}
\end{tabular}
  \label{tab:auc1}
\end{center}
\end{table}

\subsection{RQ2. What are the most important features that can help to predict if a question will receive an accepted answer?}\label{subsec:rq2}
To provide an answer to RQ2, we considered the XGBoost model as a measurement tool. As indicated, the XGBoost achieved the best AUC among the other models (see Table~\ref{tab:auc1}). XGBoost trained on all proposed features and new features performed with AUC scores of 0.71 and 0.70, while prior features resulted in an AUC of 0.66. By achieving AUC scores of 0.70 and more, our proposed features exhibit an acceptable ability to classify the questions. The results indicate that the new features can affect getting accepted answers. Our new features enhanced the prior features by adding more information regarding tags, users' experiences, and the presentation quality of questions (see Table~\ref{tab:features}). As mentioned earlier, the novel tag-related features represent the topic of a question that relates to the difficulty of providing an answer to a question and having the chance to be viewed and receive answers from more users. The new users-experience-related features, especially badges, show users' experiences and skills in asking and answering. Furthermore, to have an insight into the relative importance of the proposed features, XGBoost’s feature importance can be used. Table~\ref{tab:imp} presents a comparison between the features in terms of the XGBoost's feature importance. The ranking shows all categories of the features are important. The proposed tag-related and content-related factors attained notable results. These features show presentation quality and the effects of programming topics in attracting accepted answers. User-related features, which reflect users' skills and previous activities, also resulted in high importance scores.

\begin{table*}[ht]
  \caption{Feature ranking evaluated by XGBoost feature importance}
  \label{tab:fr}
\begin{center}
  \begin{tabular}{|lc|lc|lc|}
    \hline
    \textbf{Name}&\textbf{Importance}&\textbf{Name}&\textbf{Importance}&\textbf{Name}&\textbf{Importance}\\
    \hline
Question Creation Date&394&Asking Time (Hour)&112&Had Editor&15\\
Min Tag Quality$^{\mathrm{a}}$&349&Number of Code Snippets&106&Had Commentator&14\\
Membership Duration&317&Title Word Count &99&Had Teacher&14\\
Code Snippets Len.&315&Paragraphs&56&Had Analytical&14\\
Max Problem Rate$^{\mathrm{a}}$&311&Asking Time (Week)&54&Had Popular Question&13\\
Body AVG Word&291&Had Scholar&46&Lines of Code&12\\
AVG Tag Quality$^{\mathrm{a}}$&247&Tag Count&44&Had Good Answer&12\\
Body AVG Sentence&245&Number of Links&43&Had Enthusiast&9\\
Body Word Count&242&Had Tumbleweed&32&Has Wh-Question&8\\
Owner Questions Score&242&Is it a Question (?)&28&Has List&8\\
Owner Questions&238&Had Informed&24&Had Custodian&7\\
Title AVG Word&225&Has Error&23&Had Famous Question&6\\
Max Expert Ratio$^{\mathrm{a}}$&210&Had Student&22&Had Notable Question&5\\
Max Tag Quality$^{\mathrm{a}}$&196&Had Autobiographer&20&Had Nice Answer&4\\
Owner Accepted Answers&163&Had Supporter&18&Had Yearling&3\\
Owner Answers Score&163&Has Quote&16&Had Necromancer&2\\
Owner Answers&162&Is Capital Title&16&Had Curious&2\\
Body Sentence Count&131&{}&{}&{}&{}\\
  \hline
\multicolumn{6}{l}{$^{\mathrm{a}}$Novel tag-related features}
\end{tabular}
\label{tab:imp}
\end{center}
\end{table*} 

\subsection{RQ3. Do reading hints, rules, and information about asking questions on Stack Overflow affect getting accepted answers?}\label{subsec:rq3}
\begin{figure}[htbp]
  \centering
  \includegraphics[clip, trim=1.5cm  10.3cm 1.5cm  10cm, width=.50\textwidth]{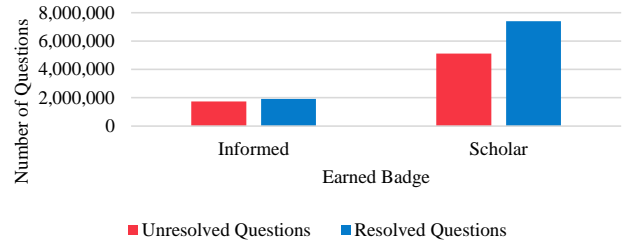}
  \caption{Numbers of unresolved and resolved questions per Scholar and Informed badges}
\label{fig:badg}
\end{figure}

Hints and documents on Stack Overflow are not significantly effective in comparison with other factors. As mentioned earlier, before and while asking a question, users read tips to ask appropriate questions. In addition, there is a badge in our feature set that explicitly indicate who have read documents on Stack Overflow. ``\verb|Informed|'' badge shows its owner has read a tour page. As shown in Table~\ref{tab:imp}, the importance of the ``\verb|Informed|'' badge was lower than the ``\verb|Scholar|'' badge, which indicates its owner previously asked a question that received an accepted answer. In another comparison, we present the number of unresolved and resolved questions for the two badges in Figure~\ref{fig:badg}. It reveals the ``\verb|Scholar|'' badge is more important than the ``\verb|Informed|'' badge. In other words, the probability of getting an accepted answer will be 0.52, if the asker is awarded with the ``\verb|Informed|'' badge, while this probability will be 0.59 if the asker earns the ``\verb|Scholar|'' badge. Furthermore, membership duration, the time from registration to asking, was an important factor. Thus, our results show that users could improve their skills more by gaining experiences than reading current documents to ask high-quality questions.

\subsection{RQ4. What roles do programming topics, tags, play in resolving questions?}\label{subsec:rq4}
\begin{figure}[htbp]
  \centering
  \includegraphics[clip, trim=1.5cm  10.7cm 1.5cm  10.2cm, width=.50\textwidth]{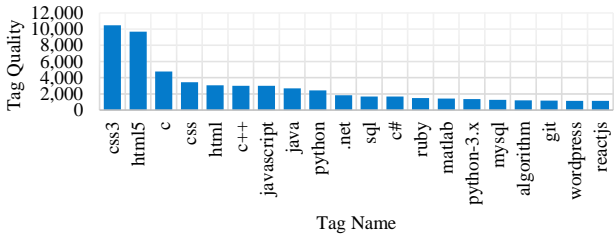}
  \caption{Top 20 best tags measured by Tag Quality}
\label{fig:tq}
\end{figure}

Tag-related features showed significant effects on getting accepted answers. The effectiveness of question answering platforms primarily depends on experts’ contributions. If a user asks a question, which is related to a topic followed by a large number of developers, it will be likely to attract more experts to answer the question and probably get an accepted answer. On the other hand, the rate of posting questions can make the distribution of experts abnormal, so we considered this in our proposed features. The importance of the novel features, which are highlighted in Table~\ref{tab:imp}, can help us to answer RQ4. The maximum of ``\verb|Problem Rate|'' reflecting the maximum value of the generating questions with a tag among 1-5 tags of a question. A low value of it logically indicates the chance of attracting followers of a tag to answer its corresponding questions. Similarly, the maximum of ``\verb|Expert Ratio|'' indicates the maximum number of possible experts who can answer  questions regarding the tags of the questions. ``\verb|Tag Quality|'' is also a reflection of attraction with respect to the {\em asking rate}. For instance, as reported in Figure~\ref{fig:tq}, ``\verb|css3|'' is the most attractive tag with regard to the {\em asking rate}.

\begin{figure}[htbp]
  \centering
  \includegraphics[clip, trim=1.5cm  10.7cm 1.5cm  10.2cm, width=.50\textwidth]{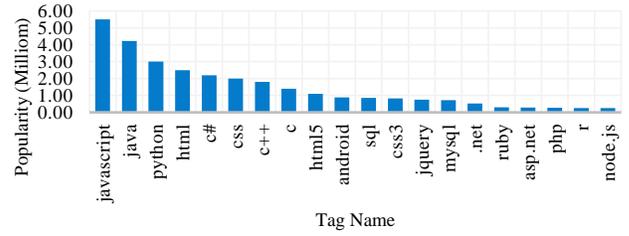}
  \caption{Top 20 best tags measured by Popularity}
  \label{fig:pop}
\end{figure}
\begin{figure}[htbp]
  \centering
  \includegraphics[clip, trim=1.5cm  10.7cm 1.5cm  10.2cm, width=.50\textwidth]{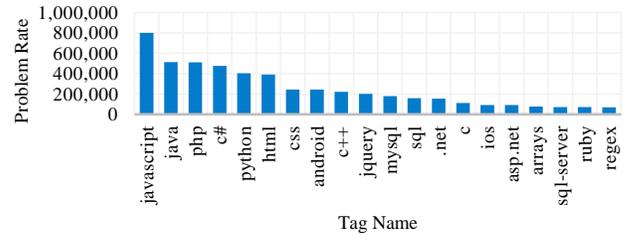}
  \caption{Top 20 best tags measured by Problem Rate}
  \label{fig:pr}
\end{figure}
\begin{figure}[htbp]
  \centering
  \includegraphics[clip, trim=1.5cm  10.7cm 1.5cm  10.2cm, width=.50\textwidth]{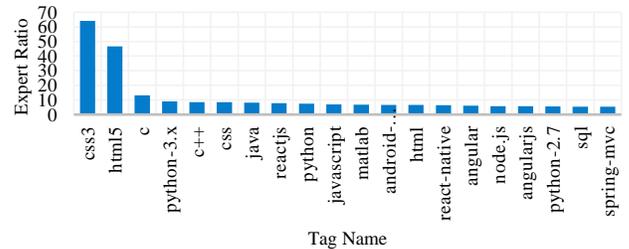}
  \caption{Top 20 best tags measured by Expert Ratio}
  \label{fig:er}
\end{figure}

By considering the proposed features, especially ``\verb|Tag Quality|'', as metrics, we can use them to compare different areas in programming, especially programming languages. In Figures \ref{fig:tq}, \ref{fig:pop}, \ref{fig:pr}, and \ref{fig:er}, we ranked top 20 tags with the frequencies of 50K or more to illustrate a comparison between popular tags.

Questions about some coding topics, especially programming language related tags, have been posted with a higher rate than others on Stack Overflow. As mentioned previously, ``\verb|Problem Rate|'' quantifies the {\em asking rate}. In the ranking of this factor (see Figure~\ref{fig:pr}), questions with ``\verb|javascript|'', ``\verb|java|'', and ``\verb|php|'' have been posted with the most rate among programming language related questions on Stack Overflow.

Concerning ``\verb|Expert Ratio|'', Figure~\ref{fig:er} shows the top 20 tags. For example, ``\verb|css3|'' attracted more users with respect to tags’ count. Furthermore, we reported a comparison between programming topics, including programming languages based on our definition of the topics’ quality in Figure~\ref{fig:tq}. As mentioned, there are flaws in the ``\verb|Problem Rate|'' and ``\verb|Expert Ratio|'' to be complete measurements in order to compare coding topics in terms of quality. For example, more popular programming languages have more tendency to generate questions, so they receive higher values of ``\verb|Problem Rate|''. Now, to overcome this problem, ``\verb|Tag Quality|'' comes into consideration. As formulated earlier, ``\verb|Tag Quality|'' represents the level of {\em asking rate} concerning its popularity. For instance, questions with ``\verb|css3|'' have high values of ``\verb|Tag Quality|'' in comparison with ``\verb|sql|''. It means, in a case, questions about ``\verb|css3|'' would be posted at the same rate as other questions about ``\verb|sql|'', but the questions with ``\verb|css3|'' tags could attract more followers at the same time. In another case, although questions about ``\verb|css3|'' and ``\verb|sql|'' attracted equal experts at the same time, but fewer questions with ``\verb|css3|'' could be asked. As a result, questions with higher values of ``\verb|Tag Quality|'' have much more chance to be resolved. This factor indicates the difficulty, attraction, and usage of a tag. Also, as presented in Figures \ref{fig:pop}, \ref{fig:pr}, and \ref{fig:er}, questions with ``\verb|javascript|'' have been submitted with the highest rate and ``\verb|javascript|'' was the most popular tag following by contributors, and questions with ``\verb|css3|'' had the ability to attract more experts in comparison with other tags.

\begin{figure}[htbp]
  \centering
  \includegraphics[clip, trim=1.5cm  9.5cm 1.5cm  9.8cm, width=.50\textwidth]{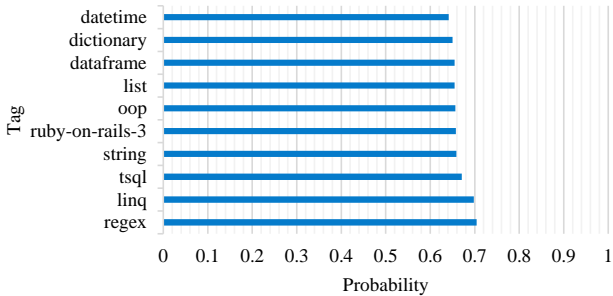}
  \caption{The probabilities of receiving accepted answers for the most probable questions’ tags}
  \label{fig:bpr}
\end{figure}
\begin{figure}[htbp]
  \centering
  \includegraphics[clip, trim=1.5cm  9.5cm 1.5cm  9.8cm, width=.50\textwidth]{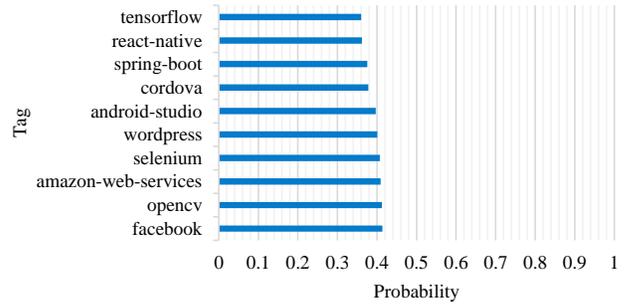}
  \caption{The probabilities of receiving accepted answers for the least probable questions’ tags}
  \label{fig:wpr}
\end{figure}

We provide a straightforward statistical report in the following. Any question probably will get an accepted answer with the probability of 52.5\%. For each tag, we calculated the probability of receiving an accepted answer for a question labeled with a specific tag. We computed the probabilities for tags that had been used in more than 1K questions. In terms of this probability, Figure~\ref{fig:bpr} shows the most probable tags to get accepted answers, whereas Figure~\ref{fig:wpr} shows the least probable tags to be resolved. The difference between them is considerable. As shown in Figures \ref{fig:bpr} and \ref{fig:wpr} indicate, general topics are more likely to be resolved than questions with technical tags. For example, a question with ``\verb|openvpn|'' has a low probability whereas another question with ``\verb|string|'' has a higher probability.

\section{Discussion}\label{sec:disc}
\begin{figure}[htbp]
  \centering
  \includegraphics[width=.45\textwidth]{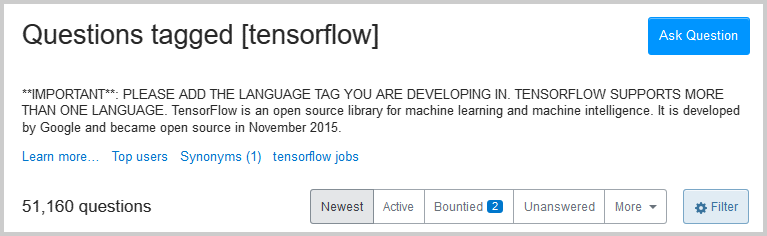}
  \caption{``tensorflow'' tag description \cite{StackExchange2020}}
  \label{fig:tens}
\end{figure}

\begin{figure}[htbp]
  \centering
  \includegraphics[clip, trim=1.4cm  11cm 1.5cm  10.4cm, width=.50\textwidth]{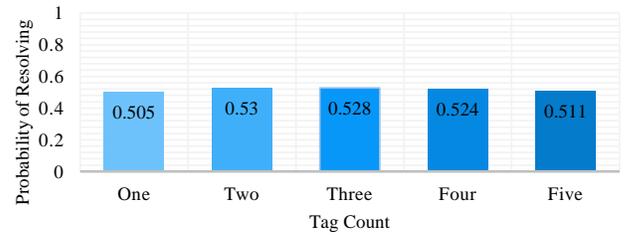}
  \caption{The probability of getting an accepted answer with respect to tag count}
  \label{fig:tc}
\end{figure}

As described, our proposed features have an adequate ability to classify questions with and without accepted answers. Here, we discuss how users can get more accepted answers in the following hints:

\begin{itemize}
\item \textbf{Content}: The proposed content-related features reflect the quality of questions. As shown in Tables \ref{tab:features} and \ref{tab:imp}, a part of important features, including the number of code snippets, the number of links, lengths, are related to the completeness, clearness, and readability of a question. The importance of these features, which reflect writing clear, complete, and readable, reveals the relationship between the quality of questions and the chance of receiving accepted answers. This report indicates clarity, readability, and completeness are important factors. This finding is compatible with Stack Overflow’s hints and documentation and Calefato et al.'s work \cite{Calefato2018}, which mentioned presentation quality. As suggested by Calefato et al. \cite{Calefato2018}, users should not ask long questions, with the lengths of 200 or more.  Concerning time to read, shorter questions are easy to answer, but removing some content from questions can cause vagueness in questions; we calculated the probability of getting an accepted answer for all questions with the lengths of 200 or more that was equal to 0.52, which is approximately the same as this probability for all questions with any lengths. Such specific rules could not be helpful enough to get more accepted answers, but our developed online tool could help askers to receive more accepted answers by checking the probable results of their questions and changing their questions based on these results.

\item \textbf{User skill}: The role of user experience in receiving accepted answers is obvious in Table~\ref{tab:imp}. The importance of users’ duration of membership, previous posts, and earned badges shows  experienced users are more successful in getting accepted answers. Due to the relationship between these features and users’ previous activities, developers can gain experiences and improve their skills in asking questions to get desired answers.

\item \textbf{Expert}: Users can increase their chances to resolve their questions by attracting more experts. The tag-related features generally contain two aspects. First, they indicate whether the topic is easy to answer or not. Second, these features reflect how many experts probably visit a question. One effective approach to attract more experts is about selecting tags. Developers can increase their chances to get accepted answers by choosing more relevant tags, but the visitors should have enough expertise. For example, if a user submits a question about IPython with both ``\verb|python|'' and ``\verb|ipython-notebook|'' instead of one of them, the user will attract experts, who can answer her/his question. In some cases, this approach is essentially important to guide questions’ visitors correctly, otherwise, experts will not visit the questions to answer them. We investigated a real example of tags on Stack Overflow. As shown in Figure~\ref{fig:wpr}, a question about ``\verb|tensorflow|'' has the lowest probability to get an accepted answer, so we searched for a possible problem with the questions tagged with ``\verb|tensorflow|''. Our observation showed among 51K questions tagged with ``\verb|tensorflow|'' only 18K of them have got accepted answers by January 2020. We found Stack Overflow has added a warning in the description of ``\verb|tensorflow|'' to inform users that they should add more tags in spite of ``\verb|tensorflow|'' (see Figure~\ref{fig:tens}). Besides, visitors, who want to post answers, should be experts in the topics. Adding tags without concerning the topic of the question is not necessarily helpful (see Figure~\ref{fig:tc}). Furthermore, the importance of asking time, in a day or a week, shows the amount of expert attraction, as suggested by Calefato et al. \cite{Calefato2018} to avoid submitting questions when the community has a lower number of visitors. On the other hand, if a large number of users try to ask at a specific time, it will cause a spike of questions in that time while an expert’s free time is limited and probably fixed.
Therefore, users should ask their questions regardless of the number of online users on Stack Overflow.
\end{itemize}


Due to the importance of user-experience-related features, we conclude that adding more appropriate documents, about asking clear, complete, readable questions, or other hints could be helpful in this regard. In fact, it can decrease the importance of user-related features, but a crucial effort is required to solve this problem; as mentioned in Section~\ref{subsec:rq3}, Stack Overflow already provided documents and hints about this, but effective lessons are required to increase users’ skills in asking better questions. Besides, users can attract more experts to get accepted answers by selecting more relevant tags. A predictive model can be beneficial to help askers in another way. As clarified, our findings did not support the guidelines regarding asking long questions or submitting questions at a specific time, but we developed an online tool to help users. They can enter their questions in this tool to find what are the probabilities of getting accepted answers for their questions. They can modify their questions and tags and check the probabilities to ask better questions.

\section{Threats to Validity}\label{sec:threats}
A threat to internal validity relates to the correctness of answers. An accepted answer that was previously correct may not be helpful anymore due to getting outdated; however, we should consider that question posers can revoke their acceptance. Also, an unaccepted answer can be a correct solution or helpful for third party users, while the asker has not accepted it. In this paper, detecting such answers was not feasible for us. Another threat to internal validity relates to closed questions. Vague, duplicate, off-topic, or opinion-based questions could be closed by moderators \cite{SO2020}. We did not filter these questions, because the reason for closing them is related to not receiving accepted answers and questions could get accepted answers before closing. Also, less than 4\% of our dataset included these questions. Threats to external validity relate to generalizability. Although we worked on 45 million questions and answers on Stack Overflow, the results of this work are not generalizable to other CQAs.

\section{Conclusion}\label{sec:conc}
Developers use Stack Overflow to solve their problems as visitors or askers. Users can ask questions in the community. Visitors also are interested in accepted answers since they can confidently reuse them, but the number of unresolved questions is increasing in the community. We analyzed 18 million questions and proposed several new features to get accepted answers. Then, to evaluate them, we built a predictive model using XGBoost with an AUC of 0.71. Finally, we identified important features to attract accepted answers and introduced an online demo tool to help developers.

\bibliographystyle{IEEEtran}
\balance
\bibliography{IEEEabrv,references}
\end{document}